\documentstyle[aps,prl,epsf,floats,multicol,amssymb,tighten]{revtex}
\begin{document}
\newcommand{\base}{}
\newcommand{\RA}{\rangle}
\newcommand{\LA}{\langle}
\newcommand{\RR}{\rangle\rangle}
\newcommand{\LL}{\langle\langle}
\newcommand{\nonb}{\nonumber}

\newenvironment{tab}[1]
{\begin{tabular}{|#1|}\hline}
{\hline\end{tabular}}

\newcommand{\fig}[2]{\epsfxsize=#1\epsfbox{#2}} \reversemarginpar 
\bibliographystyle{prsty}

\title{Slow relaxation experiments in disordered charge
and spin density waves:\\
collective dynamics of randomly distributed solitons}

\author{R. M\'elin$^{(1)}$\thanks{melin@polycnrs-gre.fr},
K. Biljakovi\'c$^{(2)}$,
J.C. Lasjaunias$^{(1)}$, and P. Monceau$^{(1)}$
}

\address{$^{(1)}$ Centre de Recherches sur les Tr\`es Basses
Temp\'eratures (CRTBT)\thanks{U.P.R. 5001 du CNRS,
Laboratoire conventionn\'e avec l'Universit\'e Joseph Fourier},\\
CNRS BP 166X, 38042 Grenoble Cedex, France}

\address{$^{(2)}$ Institute of Physics of the University,
P.O. Box 304, 41001 Zagreb, Croatia}

\date{\today}
\maketitle

\begin{abstract}
We show that the dynamics of disordered
charge density waves (CDWs) and spin density waves
(SDWs) is a collective phenomenon. The very low
temperature specific heat relaxation experiments are
characterized by:
(i) ``interrupted'' ageing (meaning
that there is a maximal relaxation time); and
(ii) a broad power-law spectrum of
relaxation times which is the signature
of a collective phenomenon. 
We propose a random energy model
that can reproduce these two observations
and from which it is possible to obtain an estimate
of the glass cross-over temperature (typically
$T_g \simeq 100 - 200$~mK).
The broad relaxation time spectrum can also be 
obtained from the solutions of two
microscopic models involving randomly distributed
solitons.
The collective behavior is similar
to domain growth dynamics in
the presence of disorder and can be described by
the dynamical renormalization group that was proposed
recently for the one dimensional random field Ising model
[D.S. Fisher, P. Le Doussal and C. Monthus,
Phys. Rev. Lett. {\bf 80}, 3539 (1998)].
The typical relaxation time scales like
$\tau^{\rm typ} \sim \tau_0 \exp{(T_g/T)}$.
The glass cross-over temperature $T_g$
related to correlations among solitons
is equal to the average energy barrier and
scales like $T_g \sim 2 x \xi_0 \Delta$. $x$
is the concentration of defects, $\xi_0$
the correlation length of the CDW or SDW
and $\Delta$ the charge or spin gap.
\end{abstract}

\widetext

\section{Introduction}

In spite of several decades of intensive experimental and
theoretical works related to slow relaxation
phenomena, important questions regarding the
nature of the low temperature phase of spin glasses have
remained unsolved. 
Replica symmetry breaking~\cite{RSB}
and droplet theory~\cite{droplet1,droplet2} constitute two 
pictures that were already available in the 80's.
It is a debated question to determine which
of these two visions of the problem does 
apply to laboratory
experiments (see for instance~\cite{Weissman}).
Recent ``memory and chaos'' experiments~\cite{Saclay}
suggest that a new type of droplet theory is needed
but there exists unsolved questions that have
been the subject of recent works~(see for
instance~\cite{Recent1,Recent2}).
Other approaches have focussed on
the description of out-of-equilibrium
ageing dynamics
in terms of generalized fluctuation-dissipation
relations~\cite{CuKu}. These ideas were originally developed
in the context of spin glass models
but have been used recently to discuss
different systems such as domain growth dynamics or
chaotic flows~\cite{Berthier}.
It has been shown recently in Ref.~\cite{Fisher-etal}
that the long time
dynamics of domain walls in the one dimensional~(1D)
random field Ising model (RFIM) could be described
by a dynamical
real space renormalization group~(RG).
The dynamical RG is
similar to the Dasgupta-Ma RG~\cite{Dasgupta-Ma}
that was applied to the  1D random
Ising model in a transverse magnetic field~\cite{Fisher92}
and to the 1D random antiferromagnetic Heisenberg
model~\cite{Fisher94}. The Dasgupta-Ma RG was also applied 
recently to many other random spin
models~\cite{Fisher92,Fisher94,Hyman96,Igloi97,Hyman97,
Monthus97,Motrunich99,Melin-et-al}.
The general purpose
of our article is to use the dynamical RG
to describe the
long time collective dynamics of
disordered charge density waves (CDWs)
and spin density waves (SDWs).

Disordered
quasi-one dimensional (quasi 1D) systems
have known a renewed interest recently since
the discovery of
several inorganic low dimensional oxides such
as CuGeO$_3$~\cite{CuGeO3-1,CuGeO3-2,CuGeO3-3,CuGeO3-4}
(a spin-Peierls compound),
PbNi$_2$V$_2$O$_8$~\cite{Pb} (a Haldane gap compound)
and Y$_2$BaNiO$_5$~\cite{ori-Y,Batlogg,Y2-1,Y2-2,Y2-3,Y2-4}
(a Haldane gap compound).
In these compounds the magnetic sites can be 
substituted with Zn (a non magnetic ion). These
substitutions generate either 3D
antiferromagnetism at low temperature
(in the case of CuGeO$_3$ and PbNi$_2$V$_2$O$_8$)
or strong 3D antiferromagnetic correlations
(in the case of Y$_2$BaNiO$_5$). In these two
types of low dimensional spin models,
the non magnetic defects
generate spin-$1/2$ moments out of the
non magnetic ground state, either in the form of
solitons in spin-Peierls
systems~\cite{Khomskii,Grenier,Saint-Paul,Dobry99,
Laukamp98,Hansen98a,Hansen98b,Augier98},
or in the form of ``edge'' spin-$1/2$ moments in 
Haldane gap systems~\cite{Kennedy,Hagiwara90,Sorensen,Yamamoto}.
The available theoretical description of these systems
is based on the Dasgupta-Ma real space
RG~\cite{Dasgupta-Ma,Bhatt-Lee}
describing non magnetic defects in 1D and
quasi-1D geometries~\cite{Y2-4,Fabrizio-Melin,Melin2}, and appears
to be in good agreement with experiments. 

It has been shown experimentally
by some of the present
authors that CDWs and SDWs present slow relaxation
phenomena at very low
temperature~\cite{manips-PF6,manips-AsF6,manips-TaS3}
under the form
of what has been called ``interrupted ageing''
in the spin glass literature~\cite{trap-model}.
In fact there are two glass cross-over temperatures.
One is due to the freezing
of the CDW domains and is observed at relatively high
temperature
mainly in dielectric susceptibility experiments
and is
defined in the
usual way for glassy systems as a separation
of the so-called $\alpha$- and $\beta$-relaxation
processes (see Ref.~\cite{alpha-beta}).
We are interested here in slow relaxation
phenomena occurring at lower temperature
that were observed in temperature relaxation
experiments~\cite{manips-PF6,manips-AsF6,manips-TaS3}.
Slow relaxation in disordered CDWs and SDWs
and antiferromagnetism
in the spin-Peierls compound CuGeO$_3$ 
have the common point that in both 
cases the physics is related to solitons having a
slow dynamics in CDWs and SDWs 
(the microscopic time associated to the
reversal of the soliton can be deduced from
experiments and is
$\tau_0 \simeq 1$~s~--~see
Ref.~\cite{manips-PF6,manips-AsF6,manips-TaS3}),
and a fast dynamics in CuGeO$_3$
(no slow relaxation has been reported in
ac-susceptibility experiments~--~see
for instance~\cite{Grenier}~-- and specific
heat relaxation experiments~\cite{Cv-Cu}).
The large value of $\tau_0$ in CDWs and SDWs
is due to energy barriers associated to the
dynamics of individual solitons, not present
in CuGeO$_3$. By comparison the microscopic time
$\tau_0 \simeq 10^{-12}$~s
in spin glasses corresponds to the reversal time
of an individual spin.
We show in this article from the analysis
of quasi 1D strong pinning models that the 
dynamics of disordered CDWs and SDWs can
be interpreted as a collective dynamics
of randomly distributed solitons. This dynamics 
is similar to a domain growth dynamics
in the presence of disorder in which 
a correlation length $\xi(t)$ is
increasing with time. Larger objects
have a slower relaxation because
most of the time
the energy barrier associated to a 
correlated pair
of solitons that are close in space is larger than the
energy barriers of the individual solitons.

It has been shown in
Refs.~\cite{Larkin1,Larkin2,Braso,Larkin3,Ovchinikov}
from the analysis of a strong pinning model 
that there exist slow relaxation phenomena associated to
independent strong pinning impurities in CDWs.
It was also shown in Ref.~\cite{Ovchinikov}
that the explanation based on independent impurities
requires an artificially
large concentration of impurities (one impurity per
unit cell). It is therefore a relevant question to
reexamine the experimental data and
investigate new mechanisms responsible for slow relaxation
in disordered CDWs and SDWs.
More precisely we address the following questions:
\begin{itemize}
\item[1.] In section~\ref{sec:exp},
we reexamine the slow relaxation experiments
in CDWs and SDWs already presented in
Ref.~\cite{manips-PF6,manips-AsF6,manips-TaS3}
and find two features:
\begin{itemize}
\item[(i)] An ``interrupted ageing'' behavior
(all relaxation times are smaller than
a maximum relaxation time $\tau_{\rm max}$).

\item[(ii)] A power-law relaxation spectrum signaling the
presence of a broad distribution of relaxation times.
\end{itemize}
The strong pinning model with independent impurities
proposed in Ref.~\cite{Ovchinikov}
has been successful to explain (i) but
it does not explain (ii).
In this article we look for a model that is consistent
with~(i) and~(ii).

\item[2.] We propose in section~\ref{sec:pheno}
a phenomenological
random energy model (REM) 
that is compatible with the slow relaxation experiments. 

\item[3.] We show in section~\ref{sec:strong-pinning}
that the experiments
can be qualitatively described as being due to
a collective behavior
in a strong pinning model. A dynamical RG is used
to describe the coupling between the solitonic deformations
of the CDW at different impurities at variance
with the previous model proposed in Ref.~\cite{Ovchinikov}
where the solitonic defects are independent from each other.

\item[4.] We show in section~\ref{sec:Ising}
that the experiments can also be qualitatively
described by the disordered spin-Peierls model
proposed in Refs.~\cite{Y2-4,Fabrizio-Melin,Melin2}.
\end{itemize}

\section{Experiments}
\label{sec:exp}

\begin{figure}[thb]
\centerline{\fig{8cm}{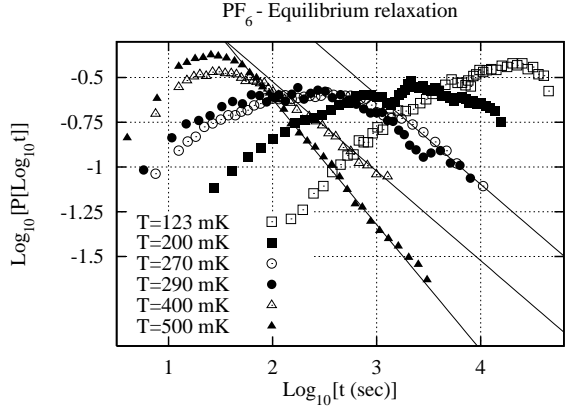}} 
\medskip
\caption{Equilibrium
relaxation time spectrum of the incommensurate
SDW compound (TMTSF)$_2$PF$_6$. Eq.~(\ref{eq:log-der}) has
been used to obtain the spectrum of relaxation
times from the temperature relaxation signal.
The waiting time
is long enough so that thermal equilibrium
has been reached for all temperatures
except $T=123$~mK.
The long-time tail of the spectrum of
relaxation times is well fitted by
a power-law:
$P_{\rm eq}(\log_{10}{t}) = 6 \times t^{-0.7}$
for $T=500$~mK;
$P_{\rm eq}(\log_{10}{t}) = 3 \times t^{-0.5}$
for $T=400$~mK;
$P_{\rm eq}(\log_{10}{t}) = 8 \times t^{-0.5}$
for $T=270$~mK. 
} 
\label{fig:PF6a}
\end{figure}
\begin{figure}[thb]
\centerline{\fig{8cm}{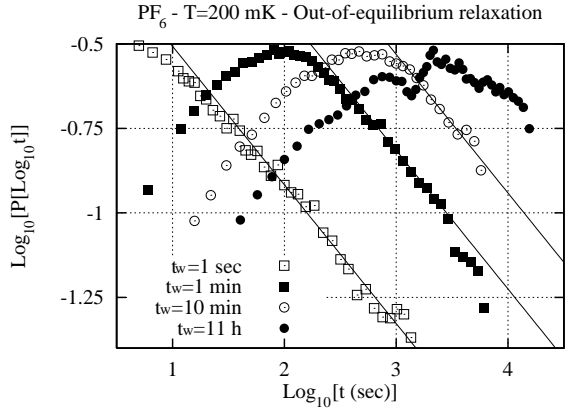}} 
\medskip
\caption{Out-of-equilibrium
relaxation time spectrum of the
incommensurate SDW compound (TMTSF)$_2$PF$_6$ at the temperature
$T=200$~mK. Eq.~(\ref{eq:log-der}) has
been used to obtain the spectrum of relaxation
times from the temperature relaxation signal.
Thermal equilibrium
has been reached for the longest relaxation
time ($t_w=11$~h).
The long-time tail of the spectrum of
relaxation times is well fitted by
a power-law:
$P_{t_w}(\log_{10}{t}) = 2.6 \times t^{-0.41}$
for $t_w=1$~sec;
$P_{t_w}(\log_{10}{t}) = 5 \times t^{-0.41}$
for $t_w=1$~min;
$P_{t_w}(\log_{10}{t}) = 0.8 \times t^{-0.41}$
for $t_w=10$~min.
} 
\label{fig:PF6b}
\end{figure}
\begin{figure}[thb]
\centerline{\fig{8cm}{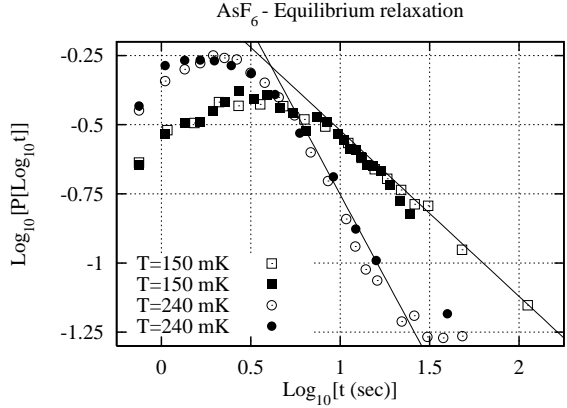}} 
\medskip
\caption{Equilibrium
relaxation time spectrum of the incommensurate
SDW compound AsF$_6$ 
(a) for $T=150$~mK and
$t_w=10$~min~($\Box$),
$t_w=1$~h~($\blacksquare$);
(b) for $T=240$~mK~ and
$t_w=10$~min~($\circ$),
$t_w=1$~h~($\bullet$).
Eq.~(\ref{eq:log-der}) has
been used to obtain the spectrum of relaxation
times from the temperature relaxation signal.
The long-time tail of the spectrum 
is well fitted by a power-law:
$P_{\rm eq}(\log_{10}{t}) = 1.2 \times t^{-0.6}$
for $T=150$~mK;
$P_{\rm eq}(\log_{10}{t}) = 2.8 \times t^{-1.2}$
for $T=240$~mK.
} 
\label{fig:AsF6}
\end{figure}
\begin{figure}[thb]
\centerline{\fig{8cm}{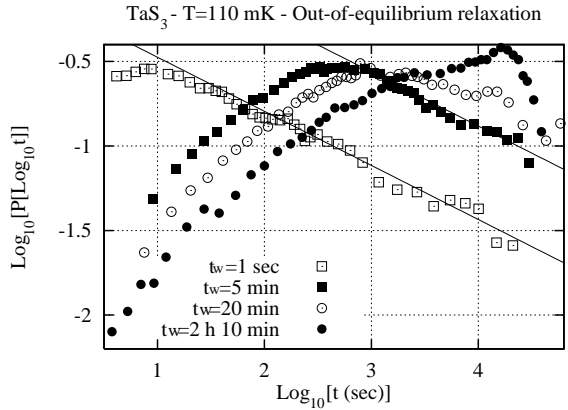}} 
\medskip
\caption{Out-of-equilibrium
relaxation time spectrum of the incommensurate
CDW compound TaS$_3$ at the temperature $T=110$~mK.
Eq.~(\ref{eq:log-der}) has
been used to obtain the spectrum of relaxation
times from the temperature relaxation signal.
The long-time tail of the spectrum 
is well fitted by a power-law:
$P_{t_w}(\log_{10}{t}) = 0.7 \times t^{-0.32}$
for $t_w=1$~sec;
$P_{t_w}(\log_{10}{t}) = 2.5 \times t^{-0.32}$
for $t_w=5$~min.
} 
\label{fig:TaS3a}
\end{figure}

\begin{figure}[thb]
\centerline{\fig{8cm}{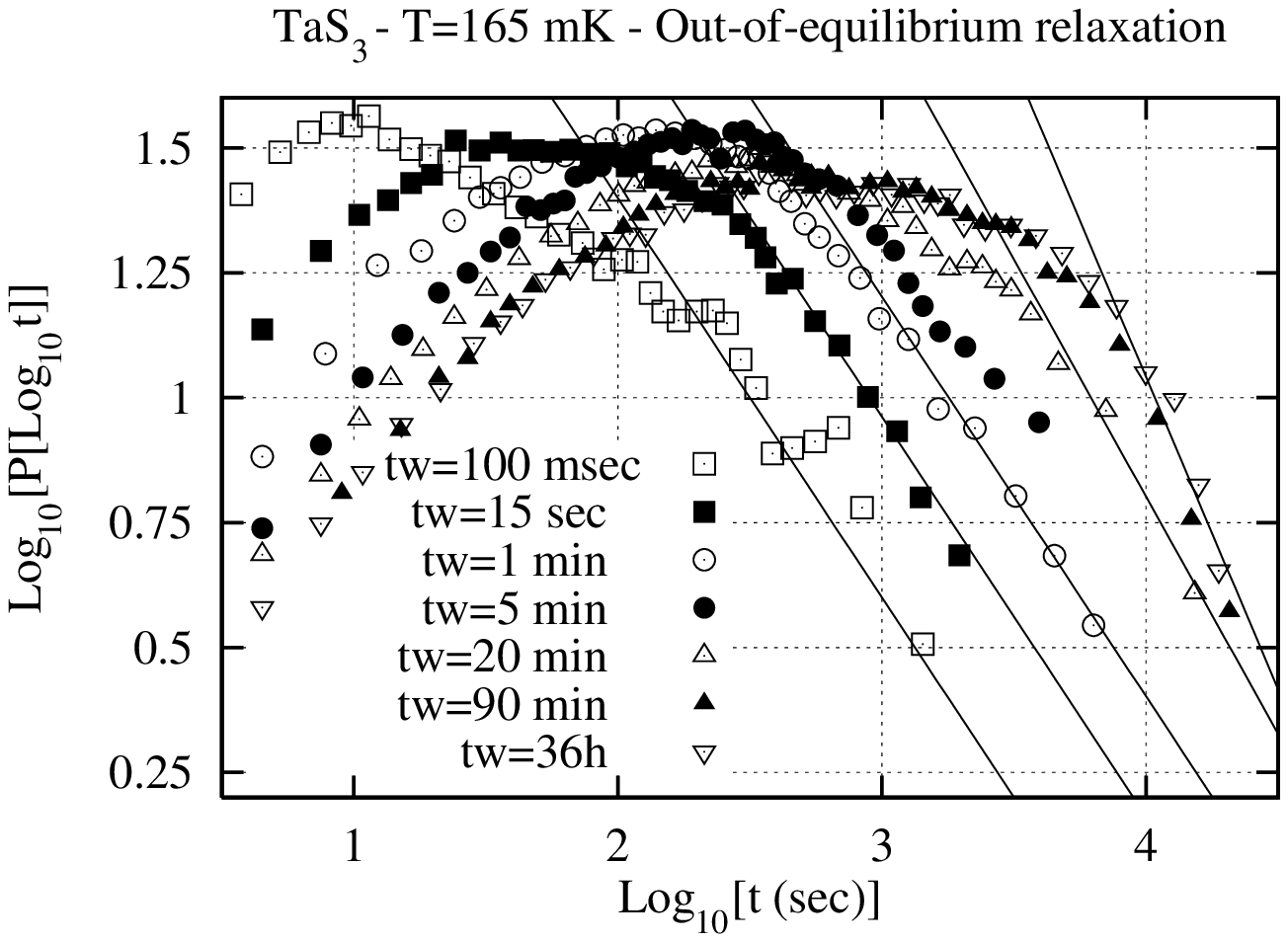}} 
\medskip
\caption{Out-of-equilibrium
relaxation time spectrum of the incommensurate
CDW compound TaS$_3$ at the temperature $T=165$~mK.
Eq.~(\ref{eq:log-der}) has
been used to obtain the spectrum of relaxation
times from the temperature relaxation signal.
The long-time tail of the spectrum 
is well fitted by a power-law:
$P_{t_w}(\log_{10}{t}) = 10^3 \times t^{-0.8}$
for $t_w=100$~msec;
$P_{t_w}(\log_{10}{t}) = 2.3 \times 10^3 \times t^{-0.8}$
for $t_w=15$~sec;
$P_{t_w}(\log_{10}{t}) = 4 \times 10^3 \times t^{-0.8}$
for $t_w=1$~min;
$P_{t_w}(\log_{10}{t}) = 4 \times 10^4 \times t^{-0.95}$
for $t_w=20$~min;
$P_{t_w}(\log_{10}{t}) = 11 \times 10^5 \times t^{-1.25}$
for $t_w=90$~min and $t_w=36$~h.
} 
\label{fig:TaS3c}
\end{figure}

The specific heat and heat relaxation
experiments~(see Refs.~\cite{manips-PF6,manips-AsF6,manips-TaS3})
have been performed at the CRTBT-Grenoble
in a dilution cryostat over the typical temperature
range $80$~mK~--~$2$~K on sample mass
of a few hundreds mg.
The experiments have been performed
with a thermal transient technique,
the sample being loosely
connected to the regulated heat sink {\sl via}
a thermal link. This technique enables us to send
energy in the sample for variable durations,
from a pulse of a fraction of seconds up to a long
``waiting time'' of $24$~h or more. For the
thermal transient experiments reported here,
the procedure is the following. Once the sample is
in equilibrium with the heat bath at $T_0$,
one increases slightly the sample
temperature to $T_0 + \Delta T_0$,
(with $\Delta T_0 / T_0\le 10 \%$) during
a waiting time $t_w$. The energy source is
switched off at $t_w$ and the thermal transient
$\Delta T(t,t_w)$ is recorded until the
temperature has relaxed to the initial
temperature $T_0$. $t$ is the time elapsed
since the waiting time $t_w$.
The temperature
relaxation $\Delta T(t,t_w)$
depends on the value of the waiting
time (ageing behavior).
If $t_w$ is sufficiently large,
$\Delta T(t,t_w)$ does not depend
on $t_w$ anymore (``interrupted
ageing'' behavior, see~\cite{trap-model}).
One can start
a new run at $T_0$ with a different
$t_w$. During a series of runs at different
$T_0$ below $1$~K, the sample is never
re-heated above $1$~--~$2$~K.
Due to the exceptional
stability of the cryostat, the reference
temperature $T_0$ can be regulated within
$\pm 2 \times 10^{-4}$ over several
tens of hours.

We use a standard procedure to deduce a spectrum
of relaxation times from the temperature relaxation
$\Delta T(t,t_w)$:
\begin{equation}
\label{eq:relax}
\Delta T(t,t_w) = 
\int  P_{t_w}(\log_{10}{\tau})
e^{-t/\tau} d \log_{10}{\tau}
.
\end{equation}
An approximate expression for the spectrum of relaxation
times can be obtained by replacing
$\exp{(-t/\tau)}$ by the $\theta$-function
$\theta( \log_{10}{\tau} - \log_{10}{t} )$,
which is justified for a broad relaxation time spectrum:
\begin{equation}
\label{eq:log-der}
P_{t_w}(\log_{10}{t}) \simeq
- \frac {\partial \Delta T(t,t_w)}
{ \partial \log_{10}{t}}
.
\end{equation}

We first consider the incommensurate SDW compound
(TMTSF)$_2$PF$_6$.
The spectrum of relaxation times $P_{t_w}(\log_{10}{t})$
obtained from Eq.~(\ref{eq:log-der})
is shown on Fig.~\ref{fig:PF6a} for
equilibrium relaxation and on Fig.~\ref{fig:PF6b}
for out-of-equilibrium relaxation.
In both cases the long time tail of the spectrum
of relaxation times is well described by
a power-law.
A similar power-law spectrum is obtained for the
incommensurate SDW compound AsF$_6$ (see Fig~\ref{fig:AsF6})
and the incommensurate CDW compound TaS$_3$
(see Figs.~\ref{fig:TaS3a} and~\ref{fig:TaS3c}).

It is visible on (TMTSF)$_2$PF$_6$ (see Fig.~\ref{fig:PF6b})
and TaS$_3$ (see Fig.~\ref{fig:TaS3c}) that
the exponent of the power-law relaxation in
the out-of-equilibrium dynamics
is independent
on the waiting time. The effect of the maximal
relation time is also visible on Fig.~\ref{fig:TaS3c}
where the long time relaxation is faster once
equilibrium has been reached.

\section{Random energy-like trap models of disordered
CDWs}
\label{sec:pheno}
\label{sec:REMtrap}
\begin{figure}[thb]
\centerline{\fig{5cm}{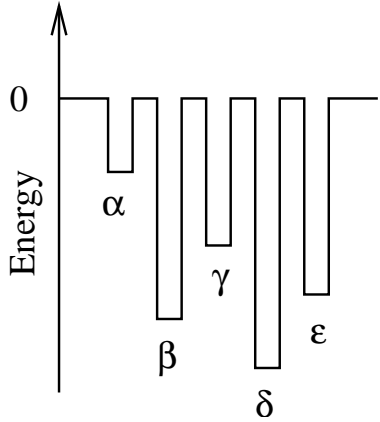}} 
\medskip
\caption{Schematic representation of
the phase space of a trap model. There
are no traps having an energy barrier larger
than $E_{\rm max}$. The maximum relaxation time
is defined as $\tau_{\rm max} = \tau_0
\exp{(E_{\rm max}/T)}$.
} 
\label{fig:schema-trap}
\end{figure}

We want to describe the slow relaxation
experiments in CDWs and SDWs discussed
in section~\ref{sec:exp} by a
REM-like trap model similar to Refs.~\cite{trap-model,Derrida}
(see Fig.~\ref{fig:schema-trap}).
The ``trap'' energies $-E_\alpha$ are
independent random variables chosen in
a distribution $p(E_\alpha)$.
The model with
$p(E_\alpha)=p_0(E_\alpha)=1/T_g \exp{(-E_\alpha/T_g)}$
solved by Bouchaud and Dean in Ref.~\cite{trap-model}
is recalled in Appendix~\ref{sec:WEB}.
We consider here a trap energy distribution
in which there is a maximal energy
barrier $E_{\rm max}$:
\begin{equation}
\label{eq:p1}
p_1(E_\alpha) = {1 \over T_g} 
\frac{\exp{(-E_\alpha/T_g)}}
{1-\exp{(-E_{\rm max}/T_g)}}
\mbox{ with $0<E_\alpha < E_{\rm max}$}
.
\end{equation}
The model with the trap distribution~(\ref{eq:p1})
has two properties related to the experiments
discussed in section~\ref{sec:exp}:
\begin{itemize}
\item[(i)] ``Interrupted ageing'' behavior due to
the presence of a maximal energy barrier. The
``complexity''
$\Omega=(T/T_g) \ln{(\tau_{\rm erg}/\tau_0)}$
introduced in Ref.~\cite{trap-model}
is equal to $\Omega=E_{\rm max}/T_g$.
\item[(ii)] A power-law distribution of relaxation times
for $\tau_\alpha < \tau_{\rm max}$:
\begin{equation}
\label{eq:p1-tau-alpha}
p_1(\tau_\alpha) = {1 \over \tau_0}
\frac{T}{T_g}
\frac{1}{1 - \exp{(-E_{\rm max})}}
\left( \frac{\tau_0}{\tau_\alpha} \right)^{1+T/T_g}
.
\end{equation}
\end{itemize}

\label{sec:REM-CDW}
\subsection{Average and typical relaxation times}
\label{sec:weak-versus-inter}

\begin{figure}[thb]
\centerline{\fig{9cm}{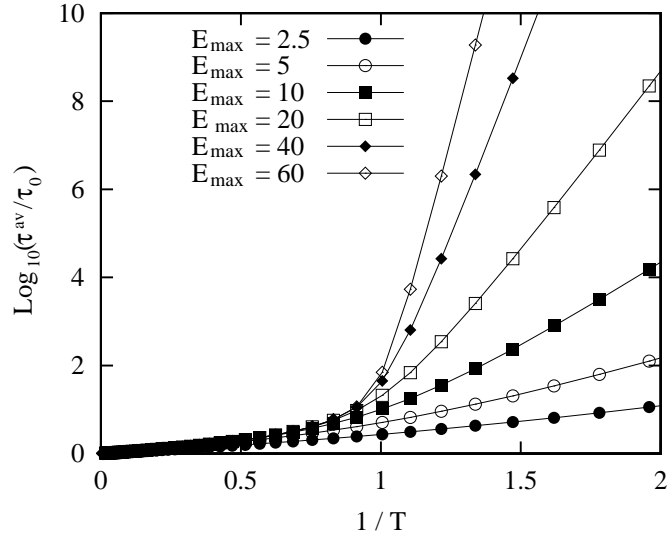}} 
\medskip
\caption{Evolution of the logarithm of the
average relaxation time
$\log_{10}{(\tau^{\rm av} / \tau_0)}$
{\sl versus} $1/T$ with $T_g=1$ and
different values of $E_{\rm max}$.
} 
\label{fig:Aharenius}
\end{figure}

The average relaxation time is
defined by
$\tau_{\rm av} = \int \tau P(\log_{10} \tau)
d \log_{10} \tau$. For the model with 
the trap energy distribution~(\ref{eq:p1}) we find
\begin{equation}
\label{eq:tau-av-REM1}
\frac{\tau^{\rm av}}{\tau^{\rm 0}}
= \frac{T}{T-T_g}
\frac{1 - \exp{\left[ - \left(1-T_g/T\right)
(E_{\rm max}/T_g) \right]}}
{1 - \exp{\left(- (E_{\rm max}/T_g) \right)}}
.
\end{equation}
The typical relaxation
time corresponds to the maximum of 
$P(\log_{10}\tau)$ and is defined as
$
\tau^{\rm typ} = \exp{ \left[ \langle \langle \ln{\tau_\alpha}
\rangle \rangle\right]}
,
$
where $\tau_\alpha$ is the trapping time in trap $\alpha$:
$\tau_\alpha=\tau_0 \exp{(E_\alpha/T)}$.
With 
the trap energy distribution~(\ref{eq:p1}) we find
\begin{equation}
\label{eq:tau-typ-REM1}
\frac{\tau^{\rm typ}}{\tau^{\rm 0}}
= \exp{\left[
\frac{T_g}{T} \left[ 1 -\frac{ (E_{\rm max}/T_g)}
{\exp{\left(E_{\rm max}/T_g\right)}-1} \right]
\right]}
,
\end{equation}
which reduces to $\tau^{\rm typ} \simeq  \tau^{\rm 0}
\exp{[T_g / T ]}$ if $E_{\rm max}$
is large compared to $T_g$. 
The activated
behavior of the typical relaxation time is in
agreement with previous experimental
observations~\cite{manips-TaS3}.
The variation of the average relaxation time
{\sl versus} $1/T$ is shown on
Fig.~\ref{fig:Aharenius}. If $E_{\rm max}$ 
is weak there is an activated behavior for all
values of the temperature, even below $T_g$.
If $E_{\rm max}$ increases there is a strong
slowing down of the dynamics below $T_g$
and the average relaxation time is infinite
when $E_{\rm max}$ is infinite in which
case we recover the behavior considered
in Ref.~\cite{trap-model}. The
average relaxation time plays a central role
in the weak ergodicity breaking scenario
for spin-glass models (see Appendix~\ref{sec:WEB})
but does not play a relevant role in 
slow relaxation in CDWs and SDWs because
it cannot be deduced from experiments
in these systems.

\subsection{Relaxation below the glass cross-over
temperature $T_g$}
 
We use the approximations presented in
Appendix~\ref{sec:WEB-approx}
to evaluate the dynamical correlation
function $\Pi_1(t,t_w)$ 
associated to
the model having a maximal energy barrier
$E_{\rm max}$:
\begin{eqnarray}
\label{eq:Pi1-below-a}
\Pi_1(t,t_w) &\simeq&
\left[1 - \left(\frac{T}{T_g}\right)^2\right]
\left( \frac{t_w}{t} \right)^{T/T_g}
\mbox{ if $t_w \ll t \ll \tau_{\rm max}$}\\
\label{eq:Pi1-below-b}
\Pi_1(t,t_w) &\simeq&
1-\left(\frac{T}{T_g}\right)^2 \left(
\frac{t}{t_w} \right)^{1-T/T_g}
\mbox{ if $t \ll t_w \ll \tau_{\rm max}$}\\
\label{eq:Pi1-below-c}
\Pi_1(t,t_w) &\simeq& 1
\mbox{ if $t \ll \tau_{\rm max} \ll t_w$}.
\end{eqnarray}
We make two comments:
\begin{itemize}
\item[(i)] The system has thermalized before $t_w$
if $t_w \gg \tau_{\rm max}$. In this case,
the correlation
function decays exponentially with time:
$\Pi(t,t_w) \sim \exp{(-t/\tau_{\rm max})}$. Within
the approximation used here, this corresponds to
$\Pi(t,t_w) \simeq 1$ if $t\ll\tau_{\rm max}$ and
$\Pi(t,t_w) \simeq 0$ if $t\gg\tau_{\rm max}$.
\item[(ii)] If $t,t_w \ll \tau_{\rm max}$, the
system has no time to ``experience'' the existence
of the maximal energy barrier $E_{\rm max}$. The
out-of-equilibrium relaxation is identical to 
the model discussed in Appendix~\ref{sec:WEB}.
\end{itemize}

\subsection{Relaxation above the
glass cross-over temperature $T_g$}
Relaxation above the
glass cross-over temperature is directly relevant to
experiments.
Above the glass cross-over temperature the correlation
function deduced from the approximations in
Appendix~\ref{sec:WEB} are found to be
\begin{eqnarray}
\label{eq:Pi1-a}
\Pi_1(t,t_w) &\simeq& 
2 \left( 1-\frac{T_g}{T} \right) \left( \frac{\tau_0}{t}
\right)^{T/T_g} \frac{t_w}{\tau_0} \mbox{ 
if $t_w \ll t \ll \tau_{\rm max}$}\\
\label{eq:Pi1-b}
\Pi_1(t,t_w) &\simeq&
\left( \frac{\tau_0}{t} \right)^{T/T_g-1}  \mbox{
if $t \ll t_w \ll \tau_{\rm max}$}\\
\label{eq:Pi1-c}
\Pi_1(t,t_w) &\simeq& 
\left( \frac{\tau_0}{t} \right)^{T/T_g-1}  \mbox{
if $t \ll \tau_{\rm max} \ll t_w$}.
\end{eqnarray}
The spectrum of relaxation times is
a power-law if $t \ll t_w \ll \tau_{\rm max}$:
$\partial \Pi_1  / \partial \ln{t}
\sim t^{1-T/T_g}$.
The exponent $1-T/T_g$ does not depend
on $t_w$. This coincides  with the experimental
behavior discussed in section~\ref{sec:exp}
(see Figs.~\ref{fig:PF6b} and~\ref{fig:TaS3a}).
The model can be used to deduce
an estimate of the glass cross-over temperature $T_g$
from the exponents obtained in experiments
(see section~\ref{sec:exp}).
$T_g$ is related to the exponent $\alpha_{\rm exp}$
appearing in the
experimental power-law relaxation spectrum 
$P_{t_w}(\log_{10}{t}) \sim t^{-\alpha_{\rm exp}}$
through the relation
$T / T_g =  1 + \alpha_{\rm exp}$.
The values of $\alpha_{\rm exp}$ and the
estimations of $T_g$ deduced from the correspondence
between the experiments and the REM-like trap model
have been given on Table~\ref{table}.
For the three compounds (TMTSF)$_2$PF$_6$,
AsF$_6$ and TaS$_3$, the exponent
$\alpha_{\rm exp}$ is increasing with
temperature which constitutes a qualitative
agreement between experiments and the
REM-like trap model. The values of $T_g$
are of the order of $T_g =
100 \div 200$~mK.
We should note that the model cannot be
used to describe heat pulse relaxation
in the regime $t \gg t_w$. Namely
Fig.~\ref{fig:PF6b} and Fig.~\ref{fig:TaS3c}
suggest that
the exponent $\alpha_{\rm exp}$ is the
same if $t \ll t_w$ or
$t_w \ll t$ while Eqs.~(\ref{eq:Pi1-a})~--~(\ref{eq:Pi1-c})
would predict a different power-law in the
regimes $t \ll t_w$ and $t_w \ll t$.
This is an indication that the exponential
trap distribution with a cut-off
is not well suited to describe the short time
dynamics.

\begin{table}
\begin{center}
\begin{tabular}{|@{}c@{}||@{}c@{}|@{}c@{}||@{}c@{}||@{}c@{}|}
\cline{1-5}
{ } Compound { } & { } $T$ (mK) { } & { } $\alpha_{\rm exp}$ { }
& { } $T/T_g$ { } & { } $T_g$ (mK){ } \\
\cline{1-5} \cline{1-5}
(TMTSF)$_2$PF$_6$ & 200 & 0.4 & 1.4 & 142 \\ 
\cline{1-5}
(TMTSF)$_2$PF$_6$ & 400 & 0.5 & 1.5 & 267 \\ 
\cline{1-5}
(TMTSF)$_2$PF$_6$ & 500 & 0.7 & 1.7 & 294 \\ 
\cline{1-5}
AsF$_6$ & 150 & 0.6 & 1.6 & 94 \\ 
\cline{1-5}
AsF$_6$ & 240 & 1.2 & 2.2 & 109 \\ 
\cline{1-5}
TaS$_3$ & 110 & 0.3 & 1.3 & 83 \\ 
\cline{1-5}
TaS$_3$ & 165 & 0.8 & 1.8 & 92 \\
\cline{1-5}
\end{tabular}
\caption{ Values of $\alpha_{\rm exp}$ 
corresponding to the experiments in section~\ref{sec:exp}.
We have indicated the estimated values of $T/T_g$ and
the estimated value of the glass cross-over
temperature $T_g$}
\label{table}
\end{center}
\end{table}

\section{Collective dynamics in a strong pinning model}
\label{sec:strong-pinning}
\label{sec:strong}
\subsection{The model}
Let us now consider the microscopic model
discussed by several authors in
Refs.~\cite{Larkin1,Larkin2,Braso,Larkin3,Ovchinikov} that
is used to describe the pinning 
of a disordered CDW:
\begin{equation}
\label{eq:H-strong}
{\cal H} =
\frac{v_F}{4 \pi} \int dx
\left( \frac{\partial \varphi(x)}{\partial x}
\right)^2 +
 w \int dx \left[ 1 - \cos{\varphi(x)} \right]
- \sum_i V_i \left[ 1 -
\cos{\left(Q x_i + \varphi(x_i) \right)} \right]
.
\end{equation}
The first term in Eq.~(\ref{eq:H-strong})
is the elastic energy, the second term
is the interchain interaction,
and the third term is the pining potential.
The charge density wave
vector is $Q$ and the impurities are at random
positions $x_i$ along the chain.

\subsection{One strong pinning impurity}
\label{sec:1imp}
Let us recall the solution of the
one-impurity model discussed
in Refs.~\cite{Larkin3,Ovchinikov}.
With $V_i=0$, $\varphi(x)$ is the solution of
the sine Gordon-like equation
\begin{equation}
\label{eq:minimum}
w \sin{\varphi(x)} - \frac{v_F}{2 \pi}
\frac{\partial^2 \varphi(x)} {\partial x^2}
- V \delta(x-x_1) \sin{\varphi(x)} =0
.
\end{equation}
The low-energy solutions are dipoles made of
a superposition of two solitons:
\begin{equation}
\label{eq:solu-1imp}
\tan{\left(\frac{\varphi(x)}{4}
\right)} = \tan{\left(
\frac{\psi}{4} \right)}
\exp{\left(-\frac{|x-x_1|}{\xi_0}\right)}
.
\end{equation}
The width of the soliton is
\begin{equation}
\label{eq:xi0}
\xi_0=\sqrt{\frac{v_F}{2 \pi w}}
.
\end{equation}
For commensurate impurities
having $Q x_1=0$, $\psi$ is the solution of
$\cos{\left(\psi/2\right)} = V_c^{(1)} / V$,
where the one-impurity pinning threshold is
given by
$V_c^{(1)} = \sqrt{ 2 w v_F / \pi}$.
With $V>V_c^{(1)}$ there is one unstable solution
($\psi=0$) and there are two stable solutions
$\psi=\pm 2 \arccos{(V_c^{(1)}/V)}$. The
single impurity solutions are degenerate and their
total energy
is $E_{\rm tot}^{(1)}=-2 (V-V_c^{(1)})^2/V$.
With incommensurate impurities ($Q x_1 \ne 0$),
$\psi$ is the solution of
\begin{equation}
\label{eq:psi-incom}
\frac{2 v_F}{\pi \xi_0} \sin{\left(\frac{\psi}{2}
\right)} = V \sin{(Q x_1 + \psi)}
.
\end{equation}
The dipolar low-energy excitations
were discussed in Ref.~\cite{Ovchinikov}
for independent impurities close to commensurability
($Q x_1 \ll 1$). 
Eq.~(\ref{eq:psi-incom}) can be solved numerically
if $Q x_1$ is not a small parameter.

Another way of solving the one-impurity model
is to consider the soliton
profile given by~(\ref{eq:solu-1imp}) as a variational
solution parametrized by $\psi$.
The energy landscape associated to a single impurity
is the following:
\begin{equation}
\label{eq:E-tot-1imp}
E_{\rm tot}^{(1)}(X_0) = 
- \frac{1}{(1+X_0^2)^2}
\left\{ A(x_1) + B(x_1) X_0
- \left( 16 w \xi_0 - C(x_1) \right) X_0^2
- B(x_1) X_0^3 - \left( 16 w \xi_0 - A(x_1) \right)
X_0^4 \right\}
,
\end{equation}
where $X_0=\tan{(\psi/4)}$.
The coefficients $A(x_1)$, $B(x_1)$ and
$C(x_1)$ are given by
\begin{eqnarray}
\label{eq:A-x1}
A(x_1) &=& 2 V \sin^2{(Q x_1)}\\
\label{eq:B-x1}
B(x_1) &=& 4 V \sin{(Q x_1)}\\
\label{eq:C-x1}
C(x_1) &=& 2 V \left[ 1 + 3 \cos{(Q x_1)} \right]
.
\end{eqnarray}
Minimizing Eq.~(\ref{eq:E-tot-1imp}) with
respect to $X_0$ leads directly to Eq.~(\ref{eq:psi-incom}).
For commensurate impurities there are two degenerate energy
minima separated by an energy barrier. For incommensurate
impurities the two energy minima are not degenerate
(there is a metastable state).

\subsection{Two strong pinning impurities}
\label{sec:2imp}

The purpose of this section is
to replace a cluster made of two impurities by
a single effective impurity. We start
with the case of two impurities
that are close to each other ($|x_2 - x_1| \ll \xi_0$).
The opposite limit $|x_2 - x_1| \gg \xi_0$ is
discussed in a straightforward fashion
because in this case the two impurities 
have an independent dynamics.
An approximate solution for an arbitrary
$|x_2-x_1|$ can be obtained by interpolating
between the two limiting cases $|x_2-x_1|
\ll \xi_0$ and $|x_2-x_1| \gg \xi_0$.

\subsubsection{Variational soliton profile}
We look for variational solutions describing two impurities at
positions $x_1$ and $x_2 > x_1$ under the form 
$\tan{ \left( \varphi(x) / 4 \right)} = F(x)$, with
\begin{equation}
\label{eq:F-test}
\label{eq:prof}
F(x) = \tan{ \left( \frac{\psi_1}{4} \right)}
\exp{\left(- \frac{|x-x_1|}{ \xi_0}\right) }
+ \tan{ \left( \frac{\psi_2}{4} \right)}
\exp{\left(- \frac{|x-x_2|}{ \xi_0} \right)}
,
\end{equation}
where $\psi_1$ and $\psi_2$ are variational parameters
that can be determined by minimizing the total energy
obtained from the soliton profile~(\ref{eq:F-test}).
If the distance between the two impurities is much
smaller than $\xi_0$ ($|x_2-x_1| \ll \xi_0$)
the two-impurity energy landscape $E_{\rm tot}^{(2)}(X_0)$
is a function of the single parameter $X_0$ given by
\begin{equation}
\label{eq:X0}
X_0 = \tan{ \left( \frac{\psi_1}{4} \right)}
+ \tan{ \left( \frac{\psi_2}{4} \right)}
.
\end{equation}

\subsubsection{Commensurate impurities in the limit
$|x_2 - x_1| \ll \xi_0$}
\label{sec:2-imp-com}

Let us now consider the situation where the
two impurities at positions $x_1$ and $x_2$ are
commensurate and the
distance between
the two impurities is small compared to the width
of the soliton ($|x_2 - x_1| \ll \xi_0$). 
The total energy of the two-impurity system takes the form
$$
E_{\rm tot}^{(2)}(X_0) =
\frac{16 X_0^2}{(1+X_0^2)^2}
\left\{ w \xi_0 - V + w \xi_0 X_0^2 \right\}
.
$$
The ground state is such that
$\partial E_{\rm tot}^{(2)}(X_0^*) / \partial X_0=0$,
which leads to
\begin{eqnarray}
X_0^* &=& \sqrt{\frac{V-w \xi_0}{V + w \xi_0}}\\
E_{\rm tot}^{(2)}(X_0^*) &=& - \frac{4}{V} \left(
V - w \xi_0 \right)^2
.
\end{eqnarray}
We make two remarks:
\begin{itemize}
\item[(i)] The pinning threshold associated to
two commensurate impurities such that
$|x_2-x_1| \ll \xi_0$ is
$V_c^{(2)} = w \xi_0=2 V_c^{(1)}$, equal to two times
the pinning threshold associated to a single impurity.
The interaction between impurities
increases the pinning threshold and
favors the so-called ``collective pinning'' regime
corresponding to $V<V_c^{(2)}$,
as opposed to the ``strong pinning'' regime
corresponding to $V>V_c^{(2)}$.

\item[(ii)] In the limit of strong impurity pinning
($V \gg w \xi_0$), there are
two energy minima corresponding to
$X_0=\tan{(\psi/4)} + \tan{(\psi'/4)} = \pm 1$ and
having an energy $E_{\rm tot}^{(2)}(X_0^*)=-4 V$.
These two minima are separated by the saddle point
$X_0'=\tan{(\psi/4)} + \tan{(\psi'/4)}=0$ having
an energy $E_{\rm tot}^{(2)}(X_0')=0$. If
$V \gg w \xi_0$ and $|x_2-x_1| \ll \xi_0$, the energy barrier
associated to the two impurity system is
$\Delta E^{(2)} = E_{\rm tot}(X_0') - E_{\rm tot}(X_0)
= 4 V$, equal to two times the energy barrier
associated to a single
impurity: $\Delta E^{(1)}=2V$ (see section~\ref{sec:1imp}).
This shows that the interaction between impurities
makes the system more glassy.
\end{itemize}

\subsubsection{Incommensurate impurities in the
limit $|x_2 - x_1| \ll \xi_0$}
\label{sec:incom}

The energy landscape associated to two incommensurate
impurities at a distance much smaller than the soliton
width~($|x_2-x_1|\ll\xi_0$)
can be reduced to an effective single impurity energy
landscape:
\begin{equation}
\label{eq:E-tot-2-gene}
E_{\rm tot}^{(2)}(X_0) = 
- \frac{1}{(1+X_0^2)^2}
\left\{ \tilde{A}_0 + \tilde{B}_0 X_0
- \left( 16 w \xi_0 - \tilde{C}_0 \right) X_0^2
- \tilde{B}_0 X_0^3 - \left( 16 w \xi_0 -
\tilde{A}_0 \right)
X_0^4 \right\}
.
\end{equation}
The coefficients $\tilde{A}_0$, $\tilde{B}_0$
and $\tilde{C}_0$ are obtained as the sum of the
coefficients associated to the
single impurity energy landscapes given by
Eqs.~(\ref{eq:A-x1})~--~(\ref{eq:C-x1}):
$\tilde{A}_0 = A(x_1) + A(x_2)$,
$\tilde{B}_0 = B(x_1) + B(x_2)$, and
$\tilde{B}_0 = C(x_1) + C(x_2)$.
The form~(\ref{eq:E-tot-2-gene}) of the
two-impurity energy landscape can be understood
as follows.
The elastic and interchain energy of the two-impurity
model are identical to the one-impurity model
if one uses the variables $X_0=\tan{(\psi/4)}$
for the one-impurity system and
$X_0=\tan{(\psi/4)} + \tan{(\psi'/4)}$
for the two-impurity system.
The pinning energy is additive
since the phase of the CDW at $x_1$
is approximately equal to the phase of the
CDW at $x_2$ if $|x_2-x_1| \ll \xi_0$.

\begin{figure}[thb]
\epsfxsize=8truecm
\begin{center}
\mbox{\epsfbox{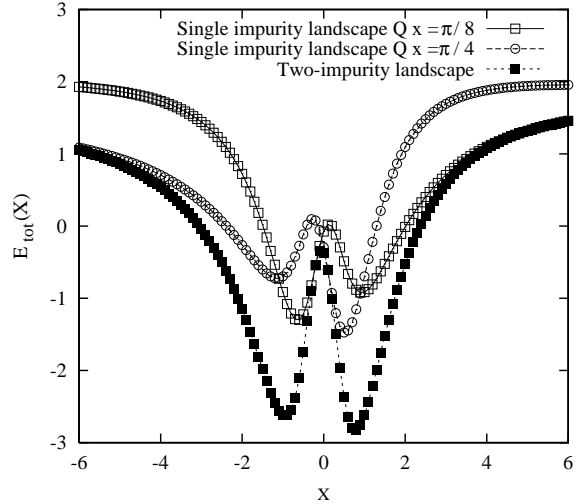}}
\end{center}
\caption{
\label{fig:landscape}
Energy landscape $E_{\rm tot}^{(1)}(X)$ {\sl versus} $X$
given by Eq.~(\ref{eq:E-tot-1imp}) of two isolated
impurities being far apart and such that
$Q x_1=\pi/8$ ($\circ$) and $Q x_2=\pi/4$ ($\Box$).
If these two impurities are at a distance much
smaller than $\xi_0$ the two-impurity system can
be represented by an effective impurity
having an
energy landscape $E_{\rm tot}^{(2)}(X)$
(see Eq.~(\ref{eq:E-tot-2-gene})).
The energy landscape $E_{\rm tot}^{(2)}(X)$
is also shown on the figure ($\blacksquare$).
}
\end{figure}

\subsubsection{Limit $|x_1 - x_2| \gg \xi_0$}

If the two impurities are at a distance much larger
than the soliton width ($|x_2-x_1| \gg \xi_0$) the
two impurities have an independent dynamics.
The two impurities are characterized by 
the coefficients $(A_1,B_1,C_1)$ and
$(A_2,B_2,C_2)$ (see Eq.~(\ref{eq:E-tot-1imp})).
The two decoupled impurities
can be replaced by a single impurity
characterized by the coefficients
$(\tilde{A}_\infty,
\tilde{B}_\infty,
\tilde{C}_\infty)$. The
coefficients $(\tilde{A}_\infty,
\tilde{B}_\infty,
\tilde{C}_\infty)$
correspond either to $(A_1,B_1,C_1)$
if the impurity at position $x_1$
has the longest relaxation time 
or to $(A_2,B_2,C_2)$ if the impurity at
position $x_2$ has the longest relaxation time.

\subsubsection{Interpolation between the
limits $|x_2 - x_1| \ll \xi_0$
and $|x_2 - x_1| \gg \xi_0$}

An arbitrary value of the distance between
the impurities can be treated by interpolating
between the two limiting cases
$|x_2 - x_1| \ll \xi_0$ and
$|x_2 - x_1 | \gg \xi_0$ discussed above.
Namely we suppose that the energy landscape
at an arbitrary distance $|x_2-x_1|$ 
is still given by Eq.~(\ref{eq:E-tot-2-gene}):
\begin{equation}
E_{\rm tot}^{(2)}(X_0) = 
- \frac{1}{(1+X_0^2)^2}
\left\{ \tilde{A} + \tilde{B} X_0
- \left( 16 w \xi_0 - \tilde{C} \right) X_0^2
- \tilde{B} X_0^3 - \left( 16 w \xi_0 - \tilde{A} \right)
X_0^4 \right\}
,
\end{equation}
and that the coefficients $\tilde{A}$, $\tilde{B}$
and $\tilde{C}$ interpolate between the
solutions already obtained in the limits
$|x_2-x_1| \ll \xi_0$ and $|x_2-x_1| \gg \xi_0$:
\begin{eqnarray}
\label{eq:inter1}
\tilde{A} = \tilde{A}_\infty
+ \left( \tilde{A}_0 - \tilde{A}_\infty \right)
\exp{\left( - \frac{ |x_2-x_1|}{\xi_0} \right)}\\
\tilde{B} = \tilde{B}_\infty
+ \left( \tilde{B}_0 - \tilde{B}_\infty \right)
\exp{\left( - \frac{ |x_2-x_1|}{\xi_0} \right)}\\
\tilde{C} = \tilde{C}_\infty
+ \left( \tilde{C}_0 - \tilde{C}_\infty \right)
\exp{\left( - \frac{ |x_2-x_1|}{\xi_0} \right)}
\label{eq:inter3}
.
\end{eqnarray}
The correlations mediated by the gaped
medium decay exponentially
with distance and this is why we use an exponential
interpolation in
Eqs.~(\ref{eq:inter1})~--~(\ref{eq:inter3}).

\subsection{Dynamical RG}
\label{sec:dynRG}
\label{sec:RG-CDW}
\begin{figure}[thb]
\centerline{\fig{8cm}{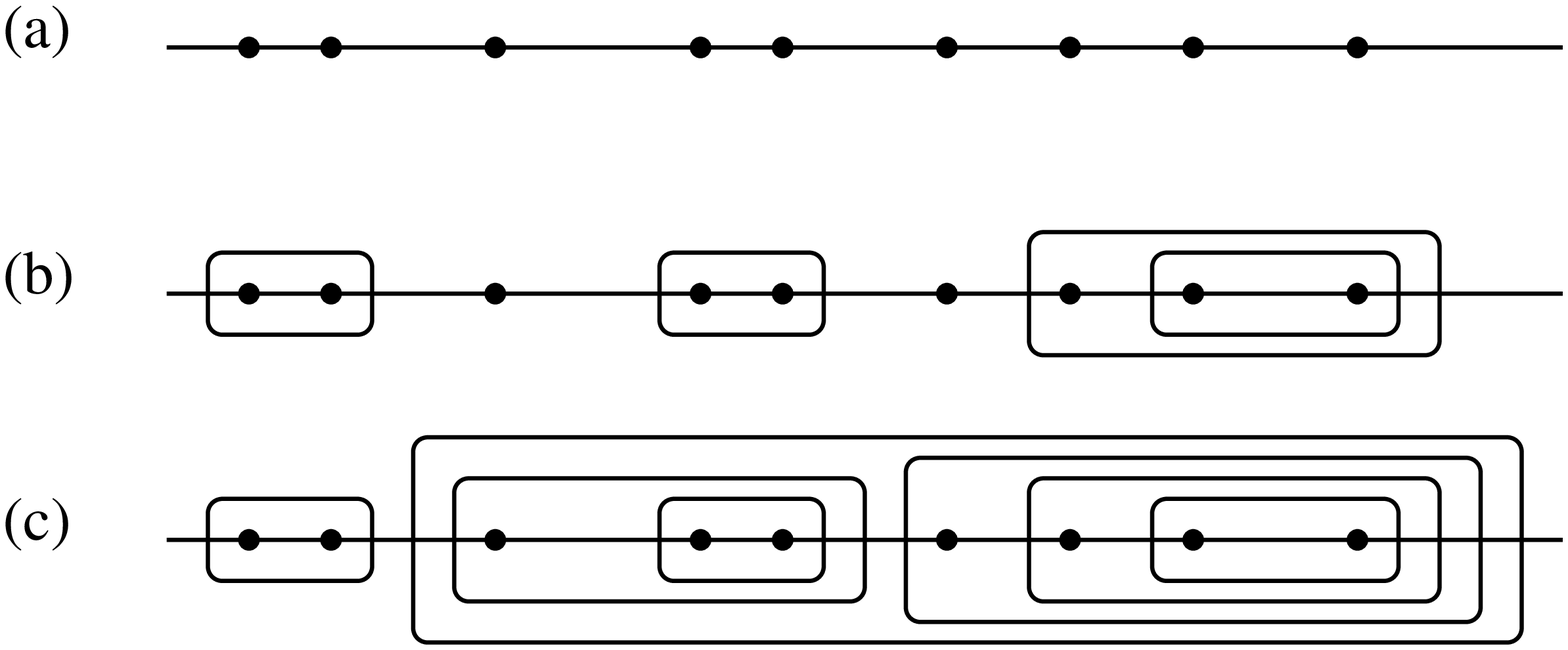}} 
\medskip
\caption{Schematic representation of the 
formation of correlated objects in the
collective dynamics.
(a) corresponds to decoupled impurities
(which is the initial condition  of the RG flow
at time $t_0$).
(b) corresponds to a time $t_1>t_0$. 
(c) corresponds to a time $t_2>t_1>t_0$.
The correlation length is increasing with
time: $\xi(t_2) > \xi(t_1) > \xi(t_0)$.
} 
\label{fig:schema}
\end{figure}

\begin{figure}[thb]
\centerline{\fig{7cm}{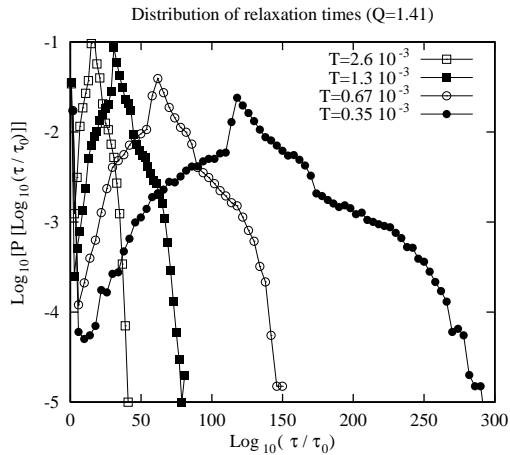}} 
\medskip
\caption{Distribution of the logarithm of the
relaxation times with an incommensurate CDW
wave vector ($Q = \sqrt{2} \simeq 1.41$).
The distribution of relaxation
times has been obtained with $N=500$ impurities
and averaged over $100$ realizations of disorder. 
The parameters of the strong pinning Hamiltonian
given by
Eqs.~(\ref{eq:H-strong}) are
$v_F=1$, $w=5 \times 10^{-6}$, $V_i=56\times 10^{-3}$
for all sites $i$.
The correlation length in the simulation
is $\xi=178 a_0$, with
$a_0$ the lattice spacing. The one-impurity
pinning threshold is $V_c=5.6 \times 10^{-3}$. The impurity
concentration is $x=0.05\ll 1$.
The temperature in units of $v_F$
corresponding to the different curves
are: $T=2.6 \times 10^{-3}$~($\Box$),
$T=1.3 \times 10^{-3}$~($\blacksquare$),
$T=0.67 \times 10^{-3}$~($\circ$),
$T=0.35 \times 10^{-3}$~($\bullet$).
The parameters are such that the width of the
soliton $\xi$ and the ratio $T/v_F$ have
the correct order of magnitude. For instance
$\xi \simeq 4000 \AA$ in CDW compounds
and $a_0 \simeq 3 \div 7 \AA$
($a_0=3.34 \AA$ in TaS$_3$ and
$a_0=7.3 \AA$ in (TMTSF)$_2$PF$_6$).
We deduce
that $\xi \simeq 500 \div 1000 a_0$ which is
compatible with the value of $\xi$ used in the simulation.
The value of the impurity concentration
$x$ is also realistic in the sense that
it has been shown experimentally that
the introduction of $0.5\%$ of extrinsic
impurities
does not modify the slow relaxation properties
so that $x$ is presumably larger than $0.5 \%$.
} 
\label{fig:distrib-incom}
\end{figure}

\begin{figure}[thb]
\centerline{\fig{7cm}{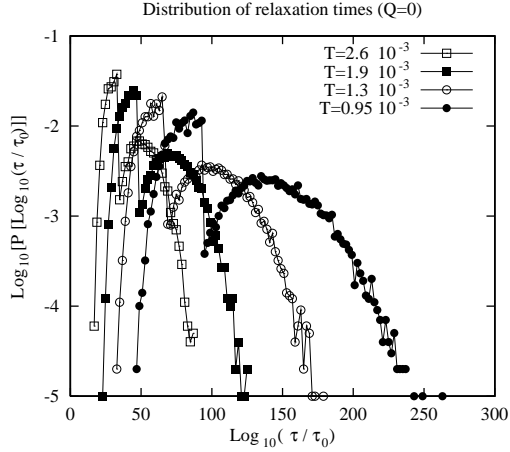}} 
\medskip
\caption{Distribution of the relaxation times
in the presence of commensurate impurities
having $Q=0$. The Hamiltonian parameters are identical
to Fig.~\ref{fig:distrib-incom}.
The temperature in units of $v_F$ are:
$T=2.6 \times 10^{-3}$~($\Box$),
$T=1.9 \times 10^{-3}$~($\blacksquare$),
$T=1.3 \times 10^{-3}$~($\circ$),
$T=0.95 \times 10^{-3}$~($\bullet$).
} 
\label{fig:distrib-com}
\end{figure}

\begin{figure}[thb]
\centerline{\fig{7cm}{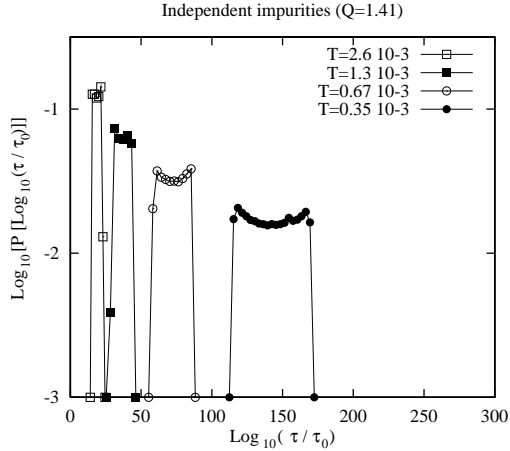}} 
\medskip
\caption{The same as Fig.~\ref{fig:distrib-incom}
but with non interacting impurities
(this is the initial condition of the RG flow).
} 
\label{fig:indep}
\end{figure}
Now we discuss the collective dynamics of the
model defined by Eq.~(\ref{eq:H-strong}).
The method is similar to
Ref.~\cite{Fisher-etal} and consists in eliminating
the fastest degrees of freedom.
There is a smallest relaxation time $\tau_{\rm min}$
(being the smallest of the relaxation
times of individual impurities) that
increases in the course of the RG.
Assuming a broad distribution of relaxation
times,
the density of relaxation times
is given by the number of impurities
$\rho(\tau_{\rm min}) \delta \tau_{\rm mnin}$
that are eliminated as the smallest relaxation time
is increased from $\tau_{\rm min}$
to $\tau_{\rm min} + \delta \tau_{\rm min}$.

We consider that the initial condition of the
dynamics is a quench from high temperature.
The initial condition of the RG flow corresponds
to  uncorrelated impurities.
The experiments discussed in section~\ref{sec:exp}
correspond to a different initial condition
but it is expected that the two types of initial
conditions can be used to discuss the qualitative
physics.

The energy landscape of
an impurity at site $x_i$
is characterized by the 
coefficients $(A(x_i),B(x_i),C(x_i))$
(see Eq.~\ref{eq:E-tot-1imp})
and by a relaxation time $\tau_i$.
We note by $\tau_{\rm min} =
{\rm Min} \{ \tau_i \}$ the smallest 
of these relaxation
times corresponding to impurity $i_0$. 
The impurity
$i_0$ has two neighboring impurities: one at the left
(site $i_L$) and one at the right (site $i_R$).
The relaxation times of the impurities at sites
$i_L$ and $i_R$ are $\tau_L$ and $\tau_R$.
There are two possibilities to eliminate
the impurity at site $i_0$:
\begin{itemize}
\item[(i)] Transform the two impurities at sites
$i_0$ and $i_L$ into
an effective impurity at site $i_L'$
having a relaxation time $\tau_L'$.
\item[(ii)] Transform the two impurities at
sites $i_0$ and $i_R$
into an effective impurity at site $i_R'$
having a relaxation time $\tau_R'$.
\end{itemize}
The transformation (i) is implemented
if $\tau_L' < \tau_R'$.
The transformation (ii) is implemented
if $\tau_R' < \tau_L'$.

The distribution of relaxation times is shown
on Fig.~\ref{fig:distrib-incom} for an incommensurate
CDW wave vector and on Fig.~\ref{fig:distrib-com}
for a commensurate CDW wave vector. The two
systems are qualitatively similar in the
sense that (i) the collective dynamics generates a broad
spectrum of relaxation times;
and (ii) there
is a maximum relaxation time $\tau_{\rm max}$.
At long times, the spectrum of relaxation times
is approximately a power-law, which is not
against the experiments
discussed in section~\ref{sec:exp}.
For comparison we have shown on Fig.~\ref{fig:indep}
the distribution of relaxation times of
non interacting strong pinning impurities
with the same parameters as Fig.~\ref{fig:distrib-incom}.
There is already a distribution of
relaxation times associated to independent impurities 
which is due to the fact that the energy landscape
associated to an impurity at position $x_1$
depends on the coordinate $x_1$ through
Eqs.~(\ref{eq:A-x1})~--~(\ref{eq:C-x1}).
But the distribution of relaxation times 
of interacting impurities
is obviously much broader than the distribution
of relaxation times of non interacting
impurities (see Figs.~\ref{fig:distrib-incom}
and~\ref{fig:indep}). 
Given that there are evidences that a broad
spectrum of relaxation times is present in
the slow relaxation experiments either for
commensurate of incommensurate systems.
we conclude that the long time dynamics of
disordered CDWs and SDWs is a collective
phenomenon that can be qualitatively
captured by the dynamical RG of the strong
pinning model. The small time
dynamics can be well approximated by 
independent impurities corresponding to
Fig.~\ref{fig:indep}
on the condition that the initial
condition is a quench from high temperature.

\section{Collective dynamics in a disordered spin-Peierls system}
\label{sec:spin-Peierls}
\label{sec:Ising}

\subsection{The model}

\begin{figure}[thb]
\centerline{\fig{11cm}{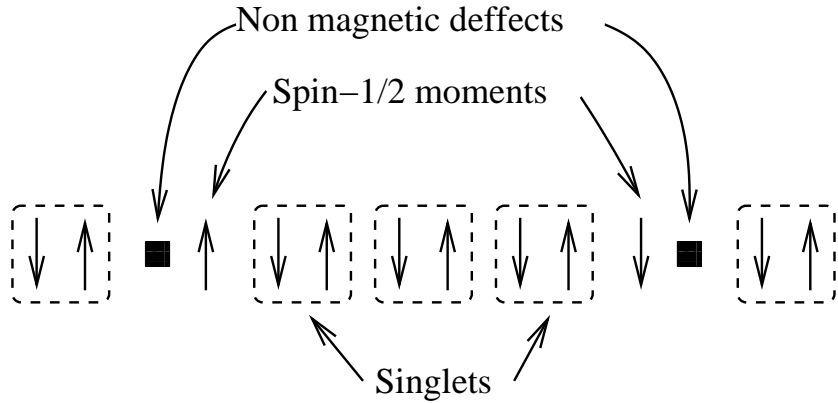}} 
\medskip
\caption{Schematic representation of 
non magnetic substitutions in a spin-Peierls system.
We assume that the
spin-$1/2$ solitonic moments have a slow dynamics.
} 
\label{fig:dim}
\end{figure}

Now we consider the model introduced
in Refs.~\cite{Y2-4,Fabrizio-Melin,Melin2} to describe
non magnetic substitutions
in spin-Peierls and Haldane gap
systems (see Fig.~\ref{fig:dim}).
In this model non magnetic impurities in a dimerized
system generate solitonic spin-$1/2$ moments out
of the non magnetic singlets. Two spin-$1/2$
moments at distance $l$ are coupled by an 
antiferromagnetic exchange
$J(l) = \Delta \exp{(-l/\xi_0)}$ that decays
exponentially with distance.
$\xi_0$ is the correlation length associated
to the gaped background.
There is an energy barrier associated
to the dynamics of an isolated soliton in 
CDWs and SDWs (see for instance
Fig.~\ref{fig:landscape}) which explains
that the microscopic time $\tau_0$ is of
order of $1$~sec
(see~\cite{manips-PF6,manips-AsF6,manips-TaS3}).
The model proposed
for non magnetic substitutions in CuGeO$_3$
can thus be ``transformed'' into a model of slow
relaxation in CDWs and SDWs just by changing
the time scale $\tau_0$ associated to the dynamics of
individual
solitonic spin-$1/2$ moments.
From this analogy we 
deduce an expression of the glass cross-over
temperature of disordered CDWs and SDWs.

We represent the solitonic spin-$1/2$ degrees
of freedom in the spin-Peierls system
by Ising spins distributed at random in 1D and
use a Glauber dynamics.
The Ising spins $\sigma_i$ interact with the
Hamiltonian
\begin{equation}
\label{eq:H-Ising}
{\cal H} = - \sum_{\langle i,j \rangle}
J_{i,j} \sigma_i \sigma_j
,
\end{equation}
where the exchange $J_{i,j}$ decays exponentially
with distance (see Refs.~\cite{Fabrizio-Melin,Melin2}):
\begin{equation}
\label{eq:H-Jij}
J_{i,j} = \Delta \exp{\left( - \frac{d_{i,j}}
{\xi_0} \right)}
.
\end{equation}
$\Delta$ is the spin gap, $\xi_0$ is the
correlation length associated to the CDW without disorder
(see Eq.~\ref{eq:xi0} for the expression of
$\xi_0$ in the strong pinning model),
and $d_{i,j}$
is the distance between the Ising spins at sites
$i$ and $j$. 
Since there is no frustration the dynamics
of the ferromagnetic
and antiferromagnetic models are equivalent
and we use here the ferromagnetic model.
\begin{figure}[thb]
\centerline{\fig{9cm}{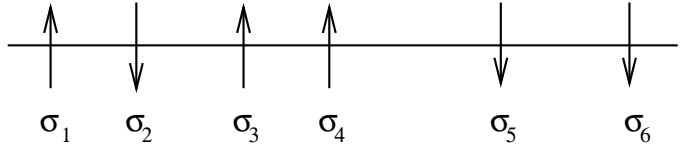}} 
\medskip
\caption{Schematic representation of
the Ising model. The spins $\sigma_i$ are
distributed at random in 1D and interact
with the Hamiltonian~(\ref{eq:H-Ising}).} 
\label{fig:trap}
\end{figure}
The single spin flip Glauber dynamics of the
model defined by Eqs.~(\ref{eq:H-Ising})
and~(\ref{eq:H-Jij}) is given by the master
equation~\cite{Glauber}
\begin{eqnarray}
\label{eq:master-eq}
\frac{d}{dt} P \left( \{ \sigma \},t \right)
= - \sum_{i=1}^N w_i \{\sigma \}
P\left( \{ \sigma \},t \right)
+ \sum_{i=1}^N
w_i \{\sigma_1, ..., - \check{\sigma}_i, ..., \sigma_N \}
P \left( \{\sigma_1, ... , - \check{\sigma}_i , ... ,
\sigma_N \},t \right)
,
\end{eqnarray}
where the transition rates take the form
\begin{equation}
\label{eq:wi}
w_i \{ \sigma \} = {r_i \over 2}
\left( 1 - \sigma_i \tanh{
\left( \beta J_{i,j} \sum_{j \in V(i)}
\sigma_j \right)} \right)
,
\end{equation}
where $V(i)$ is the set of 
neighbors of site $i$.
The form of the transition rates given by
Eq.~(\ref{eq:wi}) ensures that the
detailed balance is verified:
$P_B \{ \sigma \} w_i \{ \sigma \}
= P_B \{ \sigma' \}
w_i \{ \sigma_1 , ... , - \check{\sigma}_i
, ... , \sigma \}
,
$
where $P_B\{\sigma\}$ is the Boltzmann distribution.

\subsection{Two-spin model}
\label{sec:two-spin-model}
Following Refs.~\cite{Y2-4,Fabrizio-Melin,Melin2}
we consider the model defined by
Eqs.~(\ref{eq:H-Ising}) and~(\ref{eq:H-Jij})
for two Ising spins $\sigma_1$ and $\sigma_2$
at distance $l$. 
The distribution of distances
is $P(l) = x \exp{(-xl)}$ and the
exchange~(\ref{eq:H-Jij})
becomes $J(l) = \Delta \exp{(-l/\xi_0)}$.
The relaxation time associated to the two-spin cluster
can be obtained from the Glauber matrix defined
by~(\ref{eq:master-eq}):
$\tau(l) = \tau_0 \left[1-\tanh{(J(l)/T)} \right]^{-1}$.
Since we consider the long-time behavior, we 
use the approximation
$\tau(l) \simeq \frac{\tau_0}{2}
\exp{\left[2 J(l)/T\right]}$.
The typical relaxation time follows
an Aharenius behavior:
\begin{equation}
\label{eq:tau-typ-Ising}
\frac{\tau^{\rm typ}}{\tau_0}
= \exp{ \left(\langle \langle
\ln{\left(\frac{\tau(l)}{\tau_0} \right)}
\rangle \rangle \right)}
= \frac{1}{2}
\exp{ \left( \frac{2 x \xi_0}{1 + x \xi_0}
\frac{\Delta}{T} \right)}
.
\end{equation}
The activation energy in Eq.~(\ref{eq:tau-typ-Ising})
is equal to the average energy barrier given by
$2 \langle \langle J(l) \rangle \rangle
= 2 x \xi_0 / (1 + x \xi_0)$.
The average relaxation time is given by an
Aharenius law in limit of a small dilution of
impurities ($x \xi_0 \ll 1$):
\begin{equation}
\label{eq:tau-av-Ising}
\frac{\tau^{\rm av}}{\tau_0}
= \langle \langle \frac{\tau(l)}{\tau_0} \rangle \rangle
\simeq \frac{1}{4} 
\frac{x \xi_0 T}{\Delta}
\exp{\left(\frac{2 \Delta}{T}\right)}
.
\end{equation}
The activation energy in Eq.~(\ref{eq:tau-av-Ising})
is equal to the largest energy barrier given by
$2 {\rm Max}[ J(l) ] = 2 \Delta$.

Comparing the typical relaxation time in the
REM-like model and the two-spin model
(see Eqs.~(\ref{eq:tau-av-REM1}) and~(\ref{eq:tau-av-Ising}))
we obtain an estimate of the glass cross-over
temperature in terms of the microscopic
parameters:
\begin{equation}
\label{eq:Tg-micro}
T_g=\frac{2 x \xi_0 \Delta}{1 + x \xi_0}
.
\end{equation}
which is equal to the average energy barrier.
The estimate of $T_g$ given by Eq.~(\ref{eq:Tg-micro})
is similar to the estimate of the N\'eel temperature
of doped low dimensional oxides
discussed in Refs.~\cite{Fabrizio-Melin,Melin2}.
In both cases the physics is
controlled by correlations among
the solitons.

\subsection{RG of the disordered 1D Ising model}
\begin{figure}[thb]
\centerline{\fig{9cm}{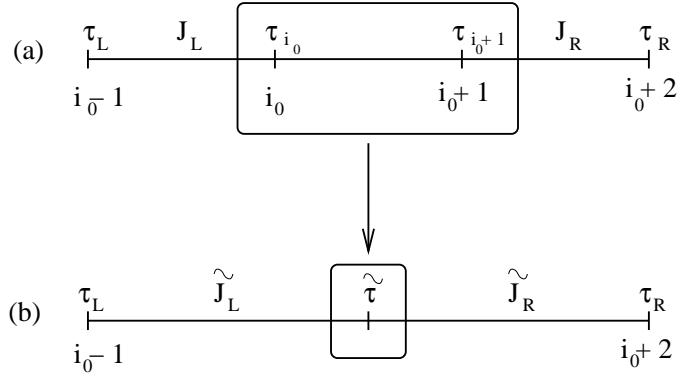}} 
\medskip
\caption{Representation of the RG
transformation. The pair of spins
$(\sigma_{i_0},\sigma_{i_0+1})$ is replaced
by the effective Ising spin $\tilde{\sigma}$
having a relaxation rate $\tilde{r}=1/\tilde{\tau}$
given by Eq.~(\ref{eq:tilder}).
} 
\label{fig:schem-Ising}
\end{figure}

Let us consider
two Ising spins $\sigma_1$ and
$\sigma_2$ coupled by a ferromagnetic exchange $J$
and replace these two spins
by an effective Ising spin $\tilde{\sigma}$.
We note $\tau_1$ and $\tau_2$ the relaxation times
of the spins $\sigma_1$ and $\sigma_2$
and we note
$r_1=1/\tau_1$ and $r_2=1/\tau_2$ the transition rates.
The Glauber matrix of the 
two-spin system can be diagonalized by forming
symmetric and antisymmetric combinations of the
occupation probabilities~\cite{Melin-Butaud}.
We deduce the transition rate $\tilde{r}$
of the effective spin $\tilde{\sigma}$:
\begin{equation}
\label{eq:tilder}
\tilde{r} = {1 \over 2}
\left\{ r_1 + r_2 - \sqrt{ (r_1 - r_2)^2
+4 r_1 r_2 \tanh^2{(\beta J)}} \right\}
.
\end{equation}
The couplings $\tilde{J}_L$ and $\tilde{J}_R$
between the effective spin $\tilde{\sigma}$
and its neighboring spins (see Fig.~\ref{fig:schem-Ising})
can be obtained by
equating the partition function of the
four-spin system  and the partition function of the
three-spin system, leading to
\begin{eqnarray}
\tilde{J}_R &=& T \arg \cosh{ \left[ \sqrt{2}
\sqrt{\cosh{(\beta J)}} \cosh{(\beta J_R)}\right]}\\
\tilde{J}_L &=& T \arg \cosh{ \left[ \sqrt{2}
\sqrt{\cosh{(\beta J)}} \cosh{(\beta J_L)}\right]}
.
\end{eqnarray}

The RG transformations can be iterated by
eliminating the smallest
relaxation time 
$\tau_{\rm min} = \mbox{Min} \{
\tau_i^{(1)} , \tau_{i,i+1}^{(2)} \}$ which
is either the relaxation time $\tau_i^{(1)}$
associated to a single spin or the relaxation
time $\tau_{i,i+1}^{(2)}$ associated to the
pair of spins $(i,i+1)$.
\begin{figure}[thb]
\centerline{\fig{9cm}{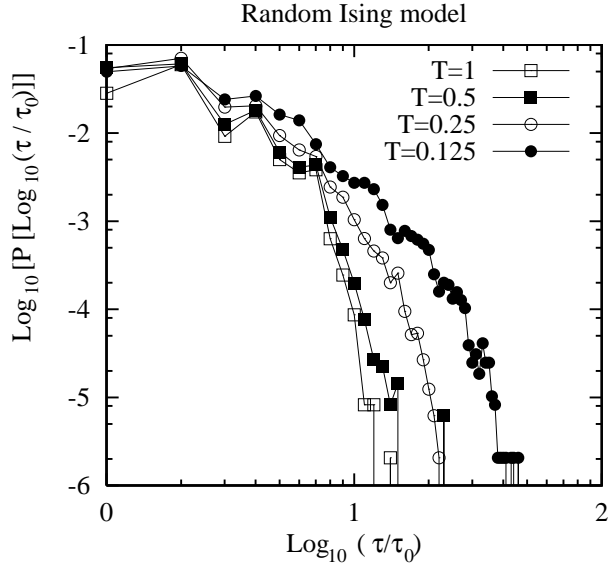}} 
\medskip
\caption{Distribution of the logarithm of the
relaxation times of the random Ising model
defined
by~(\ref{eq:H-Ising}) and~(\ref{eq:H-Jij})
with the parameters $\Delta=1$, $\xi=10$,
and $x=0.05$, with four different temperatures:
$T=1$~($\Box$),
$T=0.5$~($\blacksquare$),
$T=0.25$~($\circ$) and
$T=0.125$~($\bullet$). The ``microscopic'' time
scale is $\tau_0=1$. 
} 
\label{fig:Ising}
\end{figure}
The distribution of logarithm of the relaxation times
is shown on Fig.~\ref{fig:Ising}. Because of
the collective
dynamics there is a broad
distribution of relaxation times and a maximum
relaxation time $\tau_{\rm max}$, which is
compatible with the experiments discussed
in section~\ref{sec:exp}.
The disordered spin-Peierls system discussed in
this section has thus the same behavior as
the disordered strong pinning model discussed
in section~\ref{sec:strong}. 
In the case of the Ising model
(see Fig.~\ref{fig:Ising})
the RG generates only
time scales that are larger than $\tau_0$.
In the case of the disordered strong pinning
model the RG generates time scales
that are also smaller than $\tau_0$
(being the relaxation time of individual
solitons)
which explains the origin of the maximum
in the relaxation time spectrum
(see Fig.~\ref{fig:distrib-incom}).

\section{Conclusion}
\label{sec:conclu}

To conclude we have discussed the
collective dynamics of randomly
distributed solitons and shown that this
approach was relevant to 
disordered CDWs and SDWs.
We have proposed a phenomenological
REM-like model that in good agreement with
experiments.
Since the article was already summarized in the
Introduction, we end-up with open questions:

\begin{itemize}

\item[(i)]
We assumed that
the solitons are decoupled from
each other in the initial condition of the RG.
This corresponds to a quench
from high temperature. The initial condition
relevant to the experiments in section~\ref{sec:exp}
is different because the dynamics starts from
an equilibrium state and a small temperature variation is applied.
It has been pointed out that there can be a specific
dynamics associated to the temperature dependence of the
equilibrium correlation length~\cite{Levelut}
and it is thus a relevant question to
examine similar issues in CDWs and SDWs.

\item[(ii)] We have assumed that
only charge or spin degrees of freedom were
relevant in CDWs or SDWs. 
The interplay between charge and
spin degrees of freedom raises many questions
already in pure systems (see for instance~\cite{Poilblanc}).
It is an open question to examine the
physics associated to
defects in the presence of a charge and
a spin sector. The interplay between charge
and spin degrees of freedom could explain the
effect of a magnetic field on the slow relaxation
properties where it was observed that a weak
magnetic field has an effect on the relaxation
time spectra~\cite{H-exp}.

\end{itemize}

Finally, it would be interesting to test the
existence of slow dynamics effects in other
types of experiments, such as dielectric
response. Indeed, it is suggested in
Ref.~\cite{alpha-beta} that the residual
so-called ``$\beta_0$ process'' could
be the link to the low-$T$, slow
heat relaxation phenomena. 
Dielectric experiments in the temperature
range of $1$~K and below are planned
to test this hypothesis.

\section*{Acknowledgments}

One of us (R.M.) acknowledges fruitful
discussions on related topics with
J.C. Angl\`es d'Auriac, F. Igl\'oi and
J. Souletie.

\appendix

\section{Weak ergodicity breaking in a REM-like
trap model}
\label{sec:WEB}

In this Appendix we discuss a
REM-like trap model having
$p(E_\alpha)=p_0(E_\alpha)
={1 \over T_g} \exp{(-E_\alpha/T_g)}$
that was introduced in connection with
the so-called ``weak ergodicity breaking''
property in Ref.~\cite{trap-model}.
We first recall in section~\ref{sec:exact-solu}
the exact solution used
by Bouchaud and Dean in Ref.~\cite{trap-model}
in their discussion of the weak ergodicity
breaking property.
We give in section~\ref{sec:WEB-approx}
a set of approximations that can be used to
recover the weak ergodicity breaking property.
Similar approximations are used in 
the main body of the article
(in section~\ref{sec:pheno}) for the REM-like
trap model of disordered CDWs.
For this model relevant to disordered CDWs
it is not possible to use the same exact
solution as Bouchaud and Dean and this is
why we are lead to use a set of well controlled
approximations.

\subsection{Weak ergodicity breaking {\sl via}
an exact solution}
\label{sec:exact-solu}

The authors of Ref.~\cite{trap-model} could
demonstrate the so-called ``weak ergodicity
breaking'' property in a model
having $p(E_\alpha)=p_0(E_\alpha)
={1 \over T_g} \exp{(-E_\alpha/T_g)}$.
We note $P_\alpha(t_w)$ the probability
to find the system in trap $\alpha$ at time $t_w$.
The evolution of the system is given by the
Glauber dynamics~\cite{Glauber}
\begin{equation}
\label{eq:Glauber-trap}
\frac{d}{d t_w} P_\alpha(t_w) =
- \sum_{\beta=1,\beta \ne \alpha}^N
w_{\alpha \rightarrow \beta} P_\alpha(t_w)
+ \sum_{\beta=1,\beta \ne \alpha}^N
w_{\beta \rightarrow \alpha} P_\beta(t_w)
,
\end{equation}
where the transition rates are given by
$w_{\alpha \rightarrow \beta}=r_0
\exp{(-\beta E_\alpha)}$ and
$w_{\beta \rightarrow \alpha}=r_0
\exp{(-\beta E_\beta)}$, with
$\tau_0=1/r_0$ the microscopic time scale
and $\beta=1/T$ the inverse temperature.
To solve Eq.~(\ref{eq:Glauber-trap}), it is
convenient to make a Laplace transform
with respect to the waiting time:
$$
\tilde{P}_\alpha(E) = \int_0^{+\infty}
E d t_w  \exp{(- E t_w)} P_\alpha(t_w)
.
$$
In Laplace transform, the Glauber dynamics
equation~(\ref{eq:Glauber-trap}) becomes
\begin{equation}
\label{eq:Glauber-Laplace}
E \tilde{P}_\alpha(E) + N r_0
e^{-\beta E_\alpha} \tilde{P}_\alpha(E)
= \frac{E}{N} + \sum_\beta
r_0 e^{-\beta E_\beta}
\tilde{P}_\beta(E)
.
\end{equation}
The solution of Eq.~(\ref{eq:Glauber-Laplace})
is found to be
\begin{equation}
\label{eq:Palpha-solu}
\tilde{P}_\alpha(E) = \frac{f_E(\tau_\alpha)}
{ \sum_\beta f_E(\tau_\beta)}
,
\end{equation}
with $f_E(\tau_\alpha) = E \tau_\alpha /
(1 + E \tau_\alpha)$, and
with $\tau_\alpha=\exp{(\beta E_\alpha)}
/ (N r_0)$ the average trapping time in
trap $\alpha$. The expression of the Laplace
transform of the dynamical correlation function
is given by
\begin{equation}
\label{eq:Pi-def}
\label{eq:Laplace}
\hat{\Pi}_0(t,E) = \int_0^{+\infty}  p_0(E_\alpha)
\tilde{P}_\alpha(E) \exp{\left[-t/\tau_\alpha \right]}
d E_\alpha
.
\end{equation}
The presence of a dynamical glass transition 
can be seen from the divergence of the
average relaxation time
\begin{equation}
\label{eq:tau-av-REM-0}
\frac{\tau^{\rm av}}{\tau_0} = 
\LL \frac{\tau_\alpha}{\tau_0} \RR=
\int_0^{+\infty}
p_0(E_\alpha)
\frac{\tau_\alpha}{\tau_0}  d E_\alpha
 = \frac{T}{T-T_g}
\end{equation}
while the typical relaxation time 
\begin{equation}
\label{eq:tau-typ-REM0}
\frac{\tau^{\rm typ}}{\tau_0} =
\exp{\left[ \LL \ln{\left(
\frac{\tau_\alpha}{ \tau_0} \right)} \RR \right]}
= \exp{ \left[ {T_g \over T} \right]}
\end{equation}
follows an Aharenius behavior.
Using~(\ref{eq:Palpha-solu}),
we obtain the correlation function
\begin{equation}
\label{eq:Pi-t-E}
\hat{\Pi}_0(t,E) = \frac{
\int_{\tau_0}^{+\infty} p_0(\tau)
f_E(\tau) e^{-t/\tau} d \tau}
{ \int_{\tau_0}^{+\infty}
p_0(\tau) f_E(\tau) d \tau}
.
\end{equation}
The denominator can be evaluated
by replacing $f_E(\tau)$ by $f_0(\tau)$ defined as
$f_0(\tau) = E \tau$ if $\tau<1/E$ and
$f_0(\tau)=1$ if $\tau>1/E$. This leads to
$$
\int_{\tau_0}^{+\infty}
p_0(\tau) f_E(\tau) d \tau \simeq
\tau_0^{-T/T_g}
\frac{1}{1-T/T_g} \left\{
\frac{T_g}{T} (E \tau_0)^{T/T_g} -
E \tau_0 \right\}
.
$$
Since the waiting time is large compared to $\tau_0$,
one has $E \tau_0 \ll 1$ and therefore
$$
\int_{\tau_0}^{+\infty}
p_0(\tau) f_E(\tau) d \tau \simeq
\frac{T_g}{T(1-T/T_g)} E^{T/T_g}
.
$$
To evaluate the inverse Laplace transform
of Eq.~(\ref{eq:Pi-t-E}), Bouchaud and Dean
used the change of variable $u=f_E(\tau)$,
from what they deduced the exact result
\begin{equation}
\label{eq:Pi}
\Pi_0(t,t_w) = {T \over T_g}
\left(1-{T \over T_g}\right) \int_{\frac{t}{t+t_w}}^1 
(1-u)^{T/T_g-1} u^{-T/T_g} du
,
\end{equation}
which is valid if $T<T_g$.
In the limiting cases $t/(t+t_w) \ll 1$
and $t/(t+t_w) \simeq 1$,
Eq.~(\ref{eq:Pi-t-E}) reduces to
\begin{eqnarray}
\label{eq:exact1}
\Pi_0(t,t_w) \simeq
1 - {T\over T_g}
\left( \frac{t}{t+t_w}
\right)^{1-T/T_g} 
\mbox{if $t/(t+t_w) \ll 1$}\\
\label{eq:exact2}
\Pi_0(t,t_w) \simeq
\left(1-{T \over T_g}\right)
\left( \frac{t_w}{t+t_w} \right)^{T/T_g} 
\mbox{if $t/(t+t_w) \simeq 1$.}
\end{eqnarray}
The correlation function
$\Pi_0(t,t_w)$ tends to zero if
$t_w$ is finite and $t \rightarrow + \infty$
while it tends to unity if $t$ if finite and
$t_w \rightarrow + \infty$, which constitutes
the so-called ``weak ergodicity breaking''
property~\cite{trap-model}.

\subsection{Weak ergodicity breaking {\sl via} an
approximate solution}
\label{sec:WEB-approx}

Now let us give a set of approximations that can be used
to recover the weak ergodicity breaking property.
These approximations will be applied to another model
in section~\ref{sec:REM-CDW} in a situation where we cannot
use the change of variable leading to~(\ref{eq:Pi}).
The approximations are the following:
(i) We replace $f_E(\tau)$ by $f_0(\tau)$
in the expression~(\ref{eq:Pi-t-E}) of $\hat{\Pi}_0(t,E)$;
(ii) We replace the exponential $\exp{(-t/\tau)}$ by
the $\theta$-function $\theta (\tau-t)$.
We deduce the Laplace transform of the
dynamical correlation function
\begin{eqnarray}
\hat{\Pi}_0(t,E) \simeq 
1 - {T \over T_g} (Et)^{1-T/T_g} \mbox{ if $Et<1$}\\
\hat{\Pi}_0(t,E) \simeq 
(1-{T \over T_g}) (Et)^{-T/T_g}  \mbox{ if $Et>1$}.
\end{eqnarray}
The inverse Laplace transform is evaluated
within the same approximations. Namely we
replace $\exp{(-t/\tau_\alpha)}$ by
$\theta(\tau_\alpha-t)$. This leads to
$$
\hat{\Pi}_0(t,E) \simeq \int_0^{1/E} E d t_w
\Pi_0(t,t_w)
,
$$
from what we deduce
$$
\Pi_0(t,t_w) = \frac{\partial}
{\partial (1/E)} \left[
\frac{ \Pi_0(t,E) }{E} \right]_{E=1/t_w}
.
$$
The final form of the correlation function is found to be
\begin{eqnarray}
\label{eq:approx1}
\Pi_0(t,t_w) \simeq
1 - \left(
\frac{T}{T_g} \right)^2
\left( \frac{t}{t_w} \right)^{1-T/T_g}
\mbox{ if $t<t_w$}\\
\Pi_0(t,t_w) \simeq
\left[1-\left(\frac{T}{T_g}\right)^2\right]
\left(
\frac{t_w}{t} \right)^{T/T_g} \mbox{ if $t_w<t$}.
\label{eq:approx2}
\end{eqnarray}
The approximate correlation functions given
by~(\ref{eq:approx1})~--~(\ref{eq:approx2})
reproduce well the asymptotic behavior of
the exact solution 
given by~(\ref{eq:exact1})~--~(\ref{eq:exact2})
except for the prefactors that are
not relevant to our discussion.
The qualitative physics of the model
(namely the weak ergodicity breaking property)
can thus
be reproduced from these approximations.
We use the same approximations in the main body
of the article for the model having the trap
energy distribution~(\ref{eq:p1}) for which
we cannot use the exact solution anymore.


\end{document}